\documentclass[prl,twocolumn,amsfonts,amssymb,amsmath]{revtex4}
\usepackage{graphics}
\usepackage{epsfig}
\usepackage{psfrag}
\def\simle{\mathrel{
   \rlap{\raise 0.511ex \hbox{$<$}}{\lower 0.511ex \hbox{$\sim$}}}}
\def\simge{\mathrel{
   \rlap{\raise 0.511ex \hbox{$>$}}{\lower 0.511ex \hbox{$\sim$}}}}
\newcommand{\bq}{\begin{equation}}
\newcommand{\eq}{\end{equation}}
\def\tozero#1{\mathrel{\mathop{\sim}\limits_{\scriptscriptstyle
{#1\rightarrow0 }}}}

\newcommand{\as}{\alpha_s}

\begin{document}
\title{Geometric Scaling from DGLAP evolution}
\author{Fabrizio Caola}
\author{Stefano Forte} 
\affiliation{Dipartimento di Fisica, Universit\`a di Milano and\\ INFN, Sezione di Milano \\
Via Celoria 16, I-20133 Milano, Italy}

\begin{abstract}
We show that the geometric scaling of the total virtual photon--proton
cross section data can be explained using standard linear DGLAP perturbative
evolution  with generic boundary
conditions in a wide kinematic region. This allows us to single out
the region where geometric scaling may provide evidence for parton saturation.

\end{abstract}

\maketitle

The observation of geometric scaling~\cite{GB} in $ e p $ deep inelastic
scattering (DIS) data has attracted considerable interest because it is
widely interpreted as evidence for parton recombination and
saturation~\cite{Iancu}. Geometric scaling (GS) is the statement that the 
total $\gamma^* p$ cross section $\sigma_{tot}^{\gamma^* p}$,
which is \textit{a priori} function of two independent variables ---
the photon virtuality $Q^2$ and the Bjorken variable $x$ --- only
depends on the variable $\tau=Q^2/Q_s^2(x)$, where the so--called
saturation scale  $Q_s^2(x)$  depends nontrivially on $x$, with dimensions given
by a fixed reference scale $Q_0^2$.

The presence of recombination effects in the HERA data would have
dramatic effects, because these data dominate  the determination of 
parton distributions, which are necessary  for 
the computation of LHC
processes~\cite{Dittmar}. Available parton fits do not include these
effects and would thus fail to provide reliable predictions at the LHC.
It is the purpose of this paper to ascertain whether this is actually
the case, and more in general, in which kinematic region GS 
may or may not provide evidence for saturation.

Evidence for GS is provided by the scaling plot of the reduced cross
section $\sigma_{tot}^{\gamma^* p}$ vs. the scaling variable
$\tau$:
\bq\label{satsc}
\ln \tau=t-t_s,
\eq
where
$t=\ln Q^2/Q_0^2$, and the saturation scale $t_s\equiv \ln \frac{Q^2_s(x)}{Q_0^2}$ 
was 
originally~\cite{GB} chosen  as $t_s=\lambda \xi$, and
more recently~\cite{Iancu,Gelis} also as  $t_s=\lambda
\sqrt{\xi}$, with
$\xi=\ln(1/x)$. To test whether this behaviour is compatible with
standard DGLAP perturbative evolution, in
Figs.~\ref{teorFix}-\ref{teorRun} we show a scaling plot of
$\sigma_{tot}^{\gamma^* p}$  computed using the double asymptotic
scaling  (DAS) approximation to leading--order (LO) DGLAP
evolution~\cite{DAScit} of a constant boundary condition.
Namely, we
take
\begin{eqnarray}\label{sigmaDAS}
\sigma_{tot}^{\gamma^* p} &\equiv& \frac{4 \pi^2 \alpha_{em}}{Q^2}
F_2(x,t)\approx \frac{4 \pi^2 \alpha_{em}}{Q^2}
\frac{\gamma}{\rho}G(x,t), \\
\label{DAS}
G(\xi,t)&=&\frac{1}{\sqrt{4\pi\sigma}}
\exp\left[2\gamma \sigma-\gamma^2 \ln\left(\frac{t+\bar t_0}{\bar t_0}\right) \right], 
\end{eqnarray}
where $\sigma\equiv\sqrt{\xi\ln\frac{t+\bar t_0}{\bar t_0}}$,
$\rho\equiv\sqrt{\xi/\ln\frac{t+\bar t _0}{\bar t_0}}$, $\bar t_0\equiv\ln \frac{Q_0^2}{\Lambda_{QCD}^2}$ and 
$\beta_0=11 - \frac{2}{3}
n_f$, $\gamma=\sqrt{\frac{12}{ \beta_0}}$.

\begin{figure}
\epsfig{file=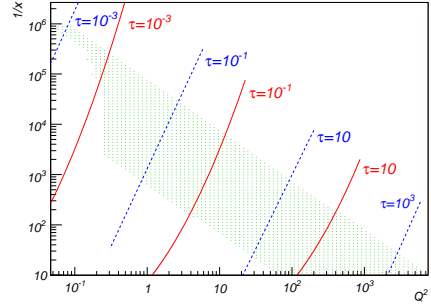,width=0.35\textwidth}
\caption{Data sample. Lines of constant $\tau$ are shown,
  with $\ln\tau=t-\lambda \xi$ (dashed) and $\ln\tau=t-\lambda \sqrt
  \xi$ (solid) and  $Q_0^2=1$~GeV$^2$.}\label{datasample}
\end{figure}

\begin{figure}
\epsfig{file=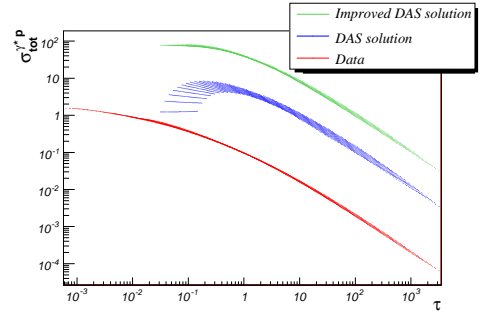,width=0.35\textwidth}
\caption{Geometric scaling for the data of Fig.~1 
with $\ln \tau=t-\lambda \xi$
  and $Q_0^2=1$~GeV$^2$. Only data with 
  $Q^2>1$~GeV$^2$ are in the DAS and improved DAS curves. The DAS
  curves are offset for clarity.}\label{teorFix}
\end{figure}

\begin{figure}
\epsfig{file=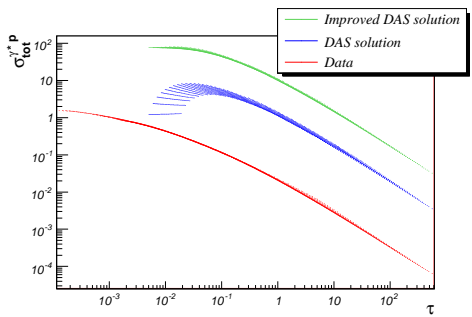,width=0.35\textwidth}
\caption{Same as Fig.~\ref{teorFix}, but  with $\ln \tau=t-\lambda
  \sqrt{\xi}$.  }\label{teorRun}
\end{figure}

The DAS approximation to 
LO DGLAP evolution is quite accurate in a wide kinematic
region $Q^2\simge10$~GeV$^2$, $x\simle0.1$, based on approximating the
LO anomalous dimension as
\bq\label{LOgamma}
\gamma_{\rm DAS}(\as,N)=\as \frac{ 3}{\pi}\left( \frac{1}{N}-1 \right).
\eq
We take it as representative of a situation where saturation is
certainly absent.
The GS properties of the cross section
Eq.~(\ref{sigmaDAS}) are compared to those of the data by plotting
both vs. $\tau$,  with the best--fit phenomenological values of
Ref.~\cite{Gelis}  $\lambda=0.321$
(Fig.~\ref{teorFix}) or
 $
\lambda=1.621$ (Fig.~\ref{teorRun}), over
 the grid of $(x,Q^2)$ values shown in Fig.~\ref{datasample}. 
 The data points are
obtained from a very accurate neural network interpolation to world
DIS data~\cite{DelDebbio}. The kinematical region is that
for which experimental data are available.

Figs.~\ref{teorFix}-\ref{teorRun} show  that for $\tau\simge1$ (Fig.~2) or
$\tau\simge0.1$ (Fig.~3)
the GS properties of the DAS
solution are almost as good as those of the data, and become as good
or better
with a minor improvement, to be discussed below. It follows
that saturation is by no means necessary for geometric
scaling. This
may seem surprizing given that the DAS solution Eq.~(\ref{sigmaDAS})
appears to violate GS. However, we now
show that approximate GS is in fact a general property of solutions to
the DGLAP equation.  It is already known that if GS is imposed as a
boundary condition at some low scale, it is preserved by both BFKL~\cite{Iancu} or
DGLAP~\cite{stasto} linear
evolution to higher scales. Here, we show instead that GS is generated by
linear DGLAP evolution itself, irrespective of the choice of boundary condition.

Consider first the fixed--coupling case.
The  general DGLAP solution for any anomalous dimension $\gamma$ is
\bq\label{genInverseMellin}
G(t,\xi)=\int_{c-i \infty}^{c+i \infty} \frac{dN}{2\pi
  i}G_0(N)\exp\left[N\xi +\gamma(\as,N)t \right],
\eq
where $G_0(N)$ is a suitable boundary condition.
For large enough  $\xi$, the integral can be evaluated  in the saddle
point approximation.   The saddle condition is
\bq\label{saddle}
\left.\frac{d}{dN}\gamma(\as,N)\right|_{N=N_0}=-\frac{\xi}{t}.
\eq
The cross section becomes
\bq\label{saddleSol}
\sigma_{tot}^{\gamma^* p}(\xi,t)\approx e^{\xi \left[N_0  +
    \left(\gamma(\as,N_0)-1\right)\frac{t}{\xi}\right]}=
\exp\left[\xi f\left(\frac{t}{\xi}\right)\right]
\eq
up to terms which are not enhanced as $\xi\to\infty$. 

Geometric
scaling follows expanding $t$ about the saturation scale
Eq.~(\ref{satsc}) 
$t_s=\lambda
\xi$:
\bq\label{Taylor}
\sigma_{tot}^{\gamma^* p}(\xi,t)\approx \exp\left[ f(\lambda)\xi + f'(\lambda)
  (t-t_s)+ \dots \right]. 
\eq
If we choose a value of $\lambda$ such that $f(\lambda)=0$ the cross
section Eq.~(\ref{Taylor}) 
manifestly displays geometric scaling. 
It is apparent from Eq.~(\ref{saddleSol}) that
this value exists if $\gamma$ is the DGLAP anomalous dimension, either
at fixed perturbative order or resummed at small $x$ using the BFKL
formalism,  or indeed for
any reasonable shape of $\gamma$.

This argument is in fact quite close to that of Ref.~\cite{Iancu},
due to the fact that  the DGLAP solution can equivalently be
written in ``dual'' form~\cite{duality} as
\bq\label{MInverseMellin}
G(t,\xi)=\int_{c-i \infty}^{c+i \infty} \frac{dM}{2\pi
  i}\bar G_0(M)\exp\left[Mt +\chi(\as,M)\xi \right],
\eq
where the kernel $\chi$ is related to $\gamma$
by
\bq\label{dual}
\chi[\as,\gamma(\as,N)]=N,
\eq 
and $\bar G_0(M)$ is determined in terms of the boundary condition
$G_0(N)$ and the anomalous dimension $\gamma$.
Evaluating the integral (\ref{MInverseMellin}) 
by saddle point and then Taylor expanding 
reproduces the argument for geometric scaling of
Ref.~\cite{Iancu}. However, Eq.~(\ref{Taylor}) shows that the
``saturation'' assumption 
of Refs.~\cite{Iancu,stasto}  that the
boundary condition satisfies GS is
redundant:  rather, GS follows from 
the existence of $\lambda$ such that  
$f(\lambda)=0$ in Eq.~(\ref{Taylor}). This is a generic
property of perturbative evolution. 
Equation~(\ref{MInverseMellin}) can be equivalently viewed as the solution
to the
BFKL or DGLAP equations, and our conclusion applies to both.

We conclude that GS holds for the solution to the DGLAP equation
at the fixed coupling level, which explains the
GS properties of the DAS solution Eq.~(\ref{sigmaDAS}),
Fig.~\ref{teorFix}: this
solution is derived with running coupling, but in practice (see 
Fig.~\ref{datasample}), the value of $t$ along fixed $\tau$ curves  
is almost constant in the data region. It follows
that $\frac{1}{\beta_0}\ln[(t+\bar t_0)/\bar t_0]\approx \as(Q_0^2)\, t$, which in
turn implies that Eq.~(\ref{DAS}) holds with 
$\sigma\approx\sqrt{\xi \beta_0\as t}$ and
$\rho\approx\sqrt{\xi/( \beta_0\as t)}$, which coincides with the
result found using Eq.~(\ref{LOgamma}) in the
approximation Eq.~(\ref{saddleSol}). 
Higher order terms in this
expansion lead to GS violations, proportional to powers of
$\as(Q_0^2)t$. The combined effect of GS violations 
will be discussed
in Fig.~4 below.

A running coupling form of GS 
can also be derived~\cite{avsar} directly for 
the cross section Eq.~(\ref{sigmaDAS}-\ref{DAS}). 
At the running coupling level, we can
neglect the variation of $\ln t/t_0$ in $\sigma$ and $\rho$ in
comparison to the scale dependence of $Q^{-2}$ in
Eq.~(\ref{sigmaDAS}). Then, the DAS solution~(\ref{DAS}) only
depends on $\sqrt{\xi}$, and the cross section~(\ref{sigmaDAS}),
consistently neglecting the variation of $\ln\xi$ in comparison to the
variation of $\xi$, is a function of
the scaling variable $t-\lambda\sqrt{\xi}$. 

Note that, unlike the fixed coupling GS in terms of $t-\lambda\xi$, 
Eq.~(\ref{Taylor}),
which holds for a generic anomalous dimension $\gamma$, this running
coupling GS depends on the particular form of the anomalous dimension
Eq.~(\ref{LOgamma}), and specifically on the fact that it has a simple
pole at $N=0$. However, this running coupling GS can also be obtained
using the running--coupling version of Eq.~(\ref{MInverseMellin})
~\cite{kancheli,dualityRun} 
\bq\label{Grun}
G(\xi,t)\approx \int\frac{dM}{2\pi
  i}\exp\Bigg[M t+ \sqrt{\xi\frac{-2
      \int_{M_0}^M\chi(\as,M')dM'}{\beta_0\as}}\Bigg],
\eq
which holds whenever the kernel $\chi$ Eq.~(\ref{dual}) is linear in
$\as$. This is the case if we only retain the simple pole in the
anomalous dimension $\gamma$ Eq.~(\ref{LOgamma}), but also for a
generic leading--order BFKL kernel, as discussed in Ref.~\cite{MP2}.
The argument leading to GS Eq.~(\ref{Taylor}) can now be repeated: 
the only difference is that the saddle point depends on
$t/\sqrt{\xi}$,
which leads to the form $t_s=\lambda\sqrt{\xi} $ of the saturation
scale. 

The arguments so far  involved several approximations. To assess their
accuracy,
we express the cross
section as a function of $\ln \tau=t-\lambda\xi$ and the orthogonal combination
$\zeta=t+\lambda\xi$. Geometric scaling is the statement that $\sigma_{tot}^{\gamma^* p}$
is independent of $\zeta$:
\bq\label{lambda}
\frac{d\sigma_{tot}^{\gamma^* p}}{d\zeta}=0.
\eq
By letting $\lambda$ depend on $\xi$
and $t$, we can view the condition ~(\ref{lambda}) as an implicit
equation for $\lambda(\xi,t)$. Geometric scaling holds to good approximation
if the solution $\lambda(\xi,t)$ to Eq.~(\ref{lambda}) 
is approximately constant in $\xi$ and $t$ in the kinematic region of interest.
\begin{figure}
\psfrag{10}{$10$}
\psfrag{e2}{$10^2$}
\psfrag{e3}{$10^3$}
\psfrag{e4}{$10^4$}
\psfrag{q2}{$Q^2$}
\psfrag{ox}{$1/x$}
\psfrag{lambda}{$\lambda$}
\epsfig{file=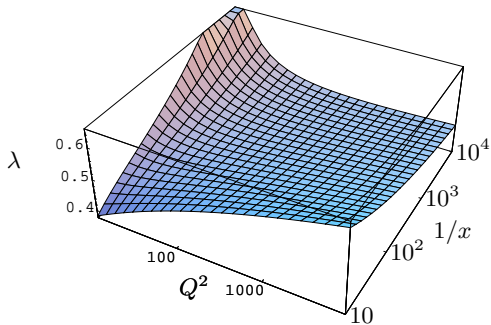,width=0.35\textwidth}
\caption{Values for $\lambda$ determined from Eq.~(\ref{lambda}).}\label{lambdaFix}
\end{figure}
\begin{figure}
\epsfig{file=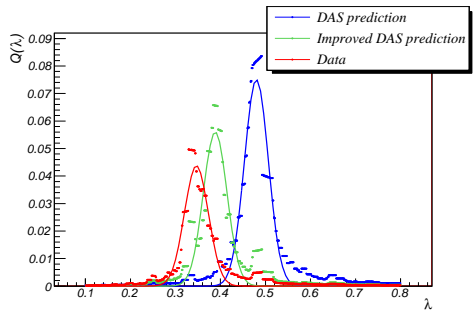,width=0.35\textwidth}
\caption{The quality factor \cite{Gelis} computed for the
  points of Fig.~\ref{teorFix} with $Q^2>25$~GeV$^2$. The solid curve
  is in each case  the result of a gaussian fit.\label{qualityfig}}
\end{figure}
We see from Fig.~\ref{lambdaFix} 
that the value of $\lambda$ is almost constant everywhere,
except at low  $Q^2\simle 25$~GeV$^2$. At  low $Q^2$, large $x$ there are no
data (see Fig.~\ref{datasample}); 
while the low $Q^2$, low $x$ region, where the DAS solution is not applicable,
shall be discussed below. We would now like to test whether
instead in the region  $Q^2\simge 25$~GeV$^2$ the approximate GS
displayed in Fig.~\ref{lambdaFix} is sufficient to explain the GS of
the data. To this purpose, we use the ``quality factor'' $Q(\lambda)$ 
which was
introduced in Ref.~\cite{Gelis} as a measure of the scaling
quality. The optimal value of $\lambda$ is that which maximizes $Q$,
and GS is better if $Q(\lambda)$ is larger.
In practice, the optimal value of $\lambda$ and the uncertainty on it
are determined by fitting a gaussian form to $Q(\lambda)$. The values
used to produce
Figs.~\ref{teorFix}-\ref{teorRun}, taken from Ref.~\cite{Gelis}, were
determined thus.

In Fig.~\ref{qualityfig} we display the quality factor $Q(\lambda)$,
computed for all points with $Q^2>25$~GeV$^2$, both for the data and
the DAS solution, as well as the result of a gaussian fit. 
We see that GS is actually rather better for the DAS
solution than for the  data. However, we also see that the optimal
value of $\lambda$ for the DAS solution is somewhat larger. This
explains why the GS properties of the DAS solution in Fig.~\ref{teorFix}
actually look a bit worse: the optimal value of $\lambda$ is
not quite the same for the data and for the DAS solution.

To explain this difference, we
note~\cite{Mankiewicz} 
that for medium-large $Q^2$ the DAS approximation can be substantially
improved by  including the contribution of
the smaller eigenvector of the anomalous dimension
matrix, whereby in Eq.~(\ref{sigmaDAS}) $F_2=(\gamma/\rho)G+\bar G$, where
$\bar G(\xi,t)=k\exp(-\delta \sigma/\rho)$, with
$\delta=16n_f/(27 \beta_0)$ and $k=0.16$ is determined from a fit to the data.
The GS plot for this improved DAS solution is also shown in
Fig.~\ref{teorFix}, and the corresponding quality factor
is displayed  in Fig.~\ref{qualityfig}: GS
  deteriorates slightly  for $Q^2>25$~GeV$^2$, but remarkably it now
  holds for all data of Fig.~\ref{teorFix}. Also, 
 the optimal value of $\lambda$ extracted from the
  data and the improved DAS solution now agree. This means that linear
  leading--order perturbative  evolution, as embodied by the
  (improved) DAS solution, can actually {\it predict} the optimal
  choice of saturation scale Eq.~(\ref{satsc}). 
Indeed, a gaussian fit
  to the quality factor for the improved DAS
  solution on all points
of Fig.~\ref{teorFix} ($Q^2>10$~GeV$^2$) gives $\lambda=0.32\pm0.05$, in perfect
  agreement with the value $\lambda=0.32\pm0.06$ determined in 
  Ref.~\cite{Gelis} from the data.  We take this as very
  strong evidence that GS for  $Q^2\simge10$~GeV$^2$ follows from
  purely linear perturbative arguments. 

A similar analysis based on Eq.~(\ref{lambda}), but with 
the running coupling form of
$\ln \tau=t-\lambda\sqrt{\xi}$ and $\zeta=t+\lambda\sqrt{\xi}$
leads to the same
  conclusion. In particular, in this case we predict $\lambda=1.66\pm 0.34$
to be compared to the experimental value $\lambda=1.62\pm 0.25$ of
Ref.~\cite{Gelis}. 

It remains to be understood why the data still display GS even at low
$Q^2$ where the DAS solution becomes unrealiable. 
 Figure~\ref{lambdaFix} suggests that
the effective value of $\lambda$ which characterizes perturbative
evolution starts growing significantly for  
$Q^2\simle 10$~GeV$^2$. 
Because GS is nevertheless seen in this data region
(see Fig.~\ref{teorFix}-\ref{teorRun}), 
one might conclude that there is some evidence for
saturation there.

However, so far we have only used   pure leading--order  DGLAP evolution,
which  fails in this region because it
does not resum small $x$ logarithms. Before concluding that
a
saturation-based approach is necessary, we should address the issue of
small $x$ resummation in the framework of linear perturbative evolution.
The small $x$ resummation of DGLAP evolution has been
recently performed, based on a suitable matching of the BFKL and DGLAP
solutions~\cite{symphen,Ciafaloni}. For our present purpose, it is
enough to consider the asymptotic small $x$
behaviour of these matched solutions, which is essentially
determined~\cite{dualityRun,symphen} by a quadratic approximation to a
running--coupling 
BFKL  evolution kernel. 

As is well known~\cite{Lipatov}, this 
leads to a solution written in terms of Airy functions
$G_A(N,t)$ if the kernel is linear in $\as$ (Bateman
functions~\cite{symphen} if the nonlinear dependence is
retained). One can then extract~\cite{dualityRun} an anomalous dimension 
\bq\label{airyad}
\gamma_A(\as(t),N)=\frac{d}{dt}\ln G_A(N,t).
\eq
The asymptotic small $x$ behaviour of DGLAP evolution at the resummed
level is controlled by the rightmost singularity of
$\gamma_A(\as(t),N)$ Eq.~(\ref{airyad}). This singularity turns out to
be a simple pole, located at $N=N_0(t)$.

Neglecting the weak~\cite{dualityRun} scale dependence of $N_0$ the
predicted asymptotic small $x$ behaviour at the resummed level is 
\bq\label{asympairy}
\sigma_{tot}^{\gamma^* p}\tozero{x} \frac{x^{- N_0}}{Q^2}.
\eq
This behaviour should hold in a region where $Q^2$ is large enough for
some resummed linear
perurbative evolution from a low--scale boundary condition to  have
taken place, say $5\simle Q^2\simle10$~GeV$^2$. In this region, we thus get
GS with $\lambda=N_0$. Typical values of $N_0$ from resummed
linear perturbative evolution are $0.1\simle N_0\simle
0.3$~\cite{symphen}. The  GS properties of the
unresummed DAS solution are thereby extended down to scales of order
$5\simle Q^2\simle10$~GeV$^2$.

The scale dependence of $N_0$ can be kept into account by determining
it  in an expansion in powers of
$\as^{2/3}(t)$~\cite{dualityRun}
\bq\label{expnzero}
N_0(t)=c\as(t)\left[1+z_0\left(\frac{\beta_0^2}{32\pi^2}
\frac{k}{c}\right)^{\frac{1}{3}} \as(t)^{\frac{2}{3}}+\dots \right],
\eq
where  $c=\frac{\chi^q(\as,M_0)}{\as}$ and
$k=\frac{1}{2\as}\frac{\partial^2}{\partial M^2}\chi^q(\as,M)|_{M=M_0} $
parametrize the quadratic BFKL kernel $\chi^q(\as,M)$ at its minimum $M=M_0$.

Substituting this in Eq.~(\ref{asympairy}) and expanding 
we see that the asymptotic
form of the cross section is constant on the curve
\bq\label{airyrun}
t(\xi)= \sqrt{\frac{4\pi c}{\beta_0}} \sqrt{\xi} +O(\xi^{1/6}).
\eq
This corresponds to the 
previously discussed running coupling form of the 
saturation scale, with $\lambda= \sqrt{\frac{4 \pi c}{\beta_0}}$.
Realistic~\cite{symphen} 
values $1\simle c\simle 2$ give  $1.2\simle \lambda\simle 1.7$. This
implies GS along this saturation line, and approximate GS in the
proximity of it. Of course, if we go very far from this region we end
up in the large $Q^2$ region which we have already discussed.

Hence,  thanks   to small $x$ resummation the growth
Fig.~\ref{lambdaFix} of $\lambda$ at small $x$ for
$5\simle Q^2\simle10$~GeV$^2$ is replaced by the (almost)  constant  value
$N_0$, thus extending GS to this region, with $t_s=N_0\xi$ or
$t_s=\sqrt{4\pi c/\beta_0}\sqrt{\xi}$.

In conclusion, we have shown that for $Q^2\simge10$~GeV$^2$ standard
linear leading-order DGLAP perturbative evolution explains geometric
scaling, and in fact predicts the value of the constant $\lambda$
which characterizes the ``saturation'' scale Eq.~(\ref{satsc}). 
Small $x$ resummation of the linear evolution equation extends
the region where GS is expected to 
values of $Q^2$ which are lower, but still within the perturbative
region. For yet lower values of $Q^2\simle5$~GeV$^2$, geometric
scaling, 
which is
observed in the data, cannot be explained using linear perturbation
theory. 
This is the region $\tau\simle 0.1$ (fixed-coupling $t_s$, Fig.~\ref{teorFix}) or 
 $\tau\simle 0.01$ (running-coupling $t_s$, Fig.~\ref{teorRun}), where the
GS plot  flattens out. In this region, geometric scaling
may provide genuine evidence for parton saturation. 

{\bf
  Acknowledgements}: We thank R.~D.~Ball and J.~Rojo for a critical
reading of the manuscript. This work was partly supported by the European
network HEPTOOLS under contract
MRTN-CT-2006-035505 and by a PRIN2006 grant (Italy).

\end{document}